\documentclass[iop]{emulateapj}
\shorttitle{An approach to account for missing redshifts}
\shortauthors{Bonaldi et al.}
\slugcomment{}
\begin{document}

\title{Cosmological constraints from Sunyaev-Zeldovich cluster counts: an approach to account for missing redshifts}

\author{A. Bonaldi, R. A. Battye and M. L. Brown}
\affil{Jodrell Bank Centre for Astrophysics, Alan Turing Building,
School of Physics and Astronomy, The University of Manchester,
Oxford Road, Manchester, M13 9PL, U.K.}
\email{anna.bonaldi@manchester.ac.uk}

\begin{abstract}

The accumulation of redshifts provides a significant observational
bottleneck when using galaxy cluster surveys to constrain cosmological
parameters. We propose a simple method to allow the use of samples
where there is a fraction of the redshifts that are not known. The
simplest assumption is that the missing redshifts are randomly extracted 
 from the catalogue, but the method also allows one to take into 
account known selection effects in the accumulation of redshifts. 
We quantify the
reduction in statistical precision of cosmological parameter constraints as a function of the fraction of
missing redshifts for simulated surveys, and also investigate the
impact of making an incorrect assumption for the distribution of
missing redshifts.

\end{abstract}


\section{Introduction}
Galaxy clusters have been used to make inferences about cosmology
using a number of different approaches \citep{Allen,Voit}. One of these is to use
the number of clusters as a function of redshift to  measure cosmological parameters, most
notably the matter density relative to critical, $\Omega_{\rm m}$,
and the amplitude of the power spectrum of density fluctuations on
small scales, $\sigma_8$. Cluster samples can be selected using the optical
richness \citep{Rozo}, X-ray luminoisity \citep{Viklinin} and the flux of
the Sunyaev-Zeldovich (SZ) effect \citep{haiman,holder,BW1}. In this paper we
will concentrate on cluster surveys selected using the SZ effect but
much of what we say can be applied to the other observables.

The SZ effect \citep{SZ} is the inverse Compton scattering of Cosmic
Microwave Background (CMB) photons as they pass through the hot
intracluster gas along the line-of-sight between the observer and the
last scattering surface of the CMB. See \cite{Birkinshaw} for a review of properties of
the SZ effect. The key feature in the context of constraining
cosmological parameters is that the SZ flux is directly proportional
to the integrated gas pressure and this is expected to be \citep{springel2001,dasilva2004,motl2005,nagai2006,kay2012} and
observed to be \citep{marrone2012,PIPSZ2013,sifon2012} an excellent proxy for the total mass of
the cluster.

Large samples of clusters selected via the SZ effect are presently being accumulated using the Atacama
Cosmology Telescope \citep[ACT,][]{ACT}, the South Pole Telescope \citep[SPT,][]{SPT2013} and the {\it Planck} satellite \citep{Planck29}. These have been used to constrain
cosmological parameters and notably in \cite{Planck20} a discrepancy
between the measured values of $\Omega_{\rm m}$ and $\sigma_8$ from
clusters and the primary CMB \citep{Planck16} has been identified. In order to 
make further progress, two things are necessary. Firstly, the understanding of the
relationship between SZ flux and the total mass of a cluster needs to
be better understood. In addition, much larger samples of clusters need
to be used to reduce the statistical uncertainties  and this requires much more
extensive knowledge of cluster redshifts. At present this creates a bootleneck in the process since each
cluster has to be targetted individually with follow-up observations. 
Large-scale surveys, such as PANSTARRS\footnote{http://pan-starrs.ifa.hawaii.edu}, DES\footnote{http://www.darkenergysurvey.org}, LSST\footnote{http://www.lsst.org} and Euclid \footnote{http://www.euclid-ec.org} might amelioarate this over significant
regions of the sky, but even then it is likely that there will not be
complete redshift coverage. 

The objective of this paper is to develop a simple method for dealing
with samples of clusters for which there is a fraction of objects (up to $20\%$,
say) with missing redshifts.  The simplest possible idea would be to
assume that the observed redshift distribution of clusters is that of
the overall sample and multiply up the observed numbers so as to get
the total number of clusters correct. However, this is too strong an
assumption. The procedure we propose is to allow clusters whose
redshifts are not known to be chosen from a statistical ensemble
depending on cosmology. Clearly this will lead to some increase
in the statistical errors bars, but this can be traded off by the
reduction in the sample size that would be necessary to obtain a pure
sample. We will discuss this in the context of simulated surveys that
are designed to be similar to those being performed by ACT, SPT and {\it
Planck}. In addition, it may be that there is some selection effect associated with the follow-up campaign, resulting in a systematic difference in the counts for known and unknown samples. If the selection effect is known then it can
easily be corrected within the proposed method. However, this may not
necessarily be the case and we discuss the impact of making an
incorrect assumption.
\section{Method}
We write the predicted number of galaxy clusters as a function of redshift for a given survey as
\begin{equation}
\frac{dN}{dz}=\int d\Omega \int \frac{dV}{dzd\Omega}\int^\infty_{M_{\rm lim}(z)} dM\frac{dn}{dM}
\end{equation}
where $d\Omega$ is the solid angle element, $dV/dzd\Omega$ is the volume element, and $dn/dM$ is the mass function. The limiting mass $M_{\rm lim}(z)$ describes the selection function of the survey. Using the SZ signal as a mass proxy, this can be related to a limit in flux density $S_{\rm lim}(z)$. A more realistic SZ selection function \citep[e.g.][]{Planck20} is expressed in terms of a limit in detection significance as a function of the size of the cluster and the position in the sky. However, such a detailed description is beyond the scope of this paper. 

By defining $N_{\rm b}$ suitable redshift bins, of width $\Delta z$, we can write the predicted number of clusters in the $i$-th bin as
\begin{equation}
N_{\rm th}(i)=\int_{zi-\Delta z/2}^{zi+\Delta z/2}dz\frac{dN}{dz},
\end{equation}
which depends on cosmology. 

The likelihood we use is

\begin{equation}
\ln P=\sum _{i=1}^{N_{\rm b}}\left[ N_{\rm cat}(i)\ln (N_{\rm th}(i))-N_{\rm th}(i)-\ln (N_{\rm cat}(i)!)\right],
\label{cash}
\end{equation}
which gives the probability of finding $N_{\rm cat}(i)$ clusters in each of $N_{\rm b}$ bins given an expected number of $N_{\rm th}(i)$, assuming they follow a Poisson distribution \citep{cash1979}. When appropriate, other likehood functions can be used to account for additional effects, such as spatial correlations between galaxy clusters \citep[e.g.][and references therein]{hu_and_cohn}. 

Computing the likelihood, being that of eq.~(\ref{cash}) or a more general one, requires that all redshifts in the catalogue are known. Let us now assume that this is not the case, i.e., the catalogue contains $N_{\rm tot}$ clusters, of which $N_{\rm red}$ have redshifts and $N_{\rm miss}$ do not. 
The minimal strategy (adopted by \citealt{Planck20}) is to multiply the distribution of known redshifts by $N_{\rm tot}/N_{\rm red}$ and use it as $N_{\rm cat}$. This assumes that the distribution of missing redshifts is exactly the same as the known ones, which would not take into account any possible statistical uncertainty. Another concern is that this strategy underestimates the error bars, which are proportional to the number of clusters in the catalogue. Both issues were negligible in \cite{Planck20}, as there was only one missing redshift out of 189 ($\sim 0.5\,\%$); however, here we want to deal with larger fractions of missing redshifts. 

An improvement over this minimal strategy would be to use the sub-sample with known $z$ to estimate the redshift distribution of the full sample by means of  resampling techniques, such as bootstrap methods \citep[see, e.g.][]{efron1982}. This strategy would capture statistical uncertainties; however, propagating the uncertainty on the resulting estimate of the full distribution to the cosmological parameters would still be an outstanding issue. The approach we advocate avoids this problem by modelling the corrections directly into the likelihood function. We modify the likelihood of eq.~(\ref{cash}) to be:
\begin{eqnarray}
\ln{P}&=&\ln \left( \frac{1}{N} \sum_{j=1}^{N} P_{j} \right) \label{cashnew}\\
\ln{P}_{j}&=&\sum _{i=1}^{N_{\rm b}}\left[N^j_{\rm cat}(i)\ln (N_{\rm th}(i))-N_{\rm th}(i)-\ln (N^j_{\rm cat}(i)!)\right] \label{cashnew2}\nonumber,
\end{eqnarray}
where $N^j_{\rm cat}$ is the $j$-th version of the catalogue counts obtained by complementing the $N_{\rm red}$ known redshifts with a random realization of the $N_{\rm miss}$ unknown redshifts.  We stress that, although in this work we are using the likelihood in eq.~(\ref{cash}), it would be straightforward to modify any likelihood function in the same way. In practice, at a given point in the cosmological parameters space we average the probability corresponding to different catalogue counts, where the missing redshifts are randomly generated. If we consider a large enough number of realizations $N$, the likelihood becomes close to deterministic. 

As a distribution for drawing the $N_{\rm miss}$ missing redshifts we can use the (normalised) theoretical counts, $N_{\rm th}(z)$, corresponding to the cosmological parameters that are currently being sampled. We note that this does not depend on the fiducial cosmology. In Sect.~\ref{sec:modelling} we introduce a modification to this simple recipe that can be used to account for systematic effects. 

We can interpret the average over the probability in eq.~(\ref{cashnew}) as a marginalisation over different versions of the catalogue. Such marginalization correctly inflates the error bars for the cosmological parameters according to the uncertainty in the redshift distribution of clusters. With respect to simply rescaling the distribution of known redshifts, we get more realistic error bars and reduce possible biases, because we are making a milder assumption on the distribution of missing redshifts. 

In this work we do not address the purity of the sample of $N_{\rm miss}$ clusters. Because false detections are typically identified with the follow-up observations, the sample of $N_{\rm miss}$ clusters could be less than 100\,\% pure. This could be accounted for statistically, by correcting $N_{\rm miss}$ for the predicted number of false detection. The purity of the sample could either be estimated with simulations or derived from that of the sample for which follow-up observations have been made. Uncertainties on the purity could be accounted for by varying $N_{\rm miss}$ according to its predicted distribution during the estimation of cosmological parameters.  

\section{Simulations}

In order to test the method we adopted the fiducial cosmology defined by the parameters of Table \ref{tab:cosmology}. We verified that this choice does not impact the results of the validation. In the following, we will denote with $N^*_{\rm th}(i)$ the predicted counts for the fiducial cosmology. We computed the constraints only for the $\Omega_{\rm m}$ and $\sigma_8$ parameters while keeping the other parameters fixed at the fiducial values. This could make our results slightly stronger than those we would expect in reality; nonetheless, the trends we observe should be representative. The estimations have been performed for 0, 5\%, 10\% and 20\% missing redshifts.

We considered two simulated surveys, as detailed below, using either a limit in flux density or a limit in mass as a selection function. 
To convert from the SZ flux, $Y_{500}$,  to cluster mass, $M_{500}$, integrated within the radius where the mean enclosed density is 500 times the critical density, we adopted a relation of the form
\begin{equation}
\resizebox{.9\hsize}{!}{$Y_{500}=Y_* \left ( \frac{M_{500}}{3 \times 10^{14}h^{-1}M_{\odot}}\right )^{\alpha} E(z)^{2/3} \left ( \frac{d_{\rm A}}{500\, {\rm Mpc}}\right )^{-2}$}\label{ymrel}
\end{equation}
where $d_{\rm A}(z)$ is the angular diameter distance to redshift $z$, $E^2(z)=\Omega_{\rm m}(1+z)^3+\Omega_\Lambda$, $Y_*=8.9\times 10^{-4}$ and $\alpha=1.79$. 
This is very similar to the one used in \cite{Planck20}.
We used the mass function by \cite{tinker}.
\begin{table}
\begin{center}
\caption{Fiducial cosmological parameters of the simulation}
\begin{tabular}{cccccl}
\hline
$\Omega_{\rm m}$&$\Omega_{\rm b}$&$\Omega_\Lambda$&$\sigma_8$&$n_{\rm s}$&$H_0$\,[Km/s/Mpc]\\
\hline
0.3&0.04&0.7&0.8&1&70\\
\hline
\end{tabular}
\label{tab:cosmology}
\end{center}
\end{table}

\begin{figure}
\includegraphics[width=5.8cm,angle=90]{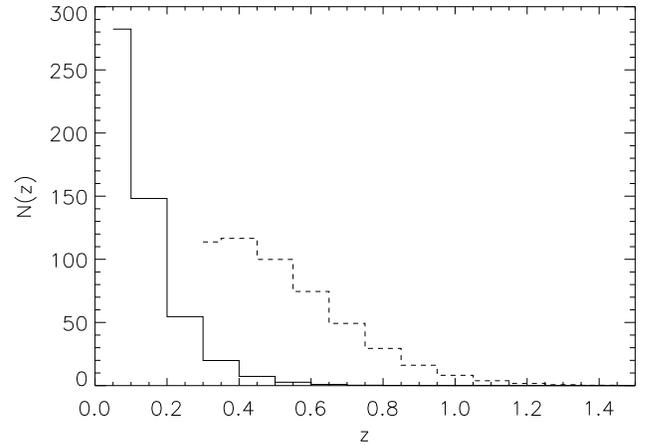}
\caption{Predicted counts in $z$ for Survey 1 (solid line) and Survey 2 (dashed line).}
\label{fig:planck_counts}
\end{figure}

\paragraph{Survey 1} The first survey attempts to mimic the catalogue expected for {\it Planck} at $S/N>5$. It has a large sky coverage (65\% of the sky, 26814.4\,deg$^2$) and a constant flux limit of $Y^{\rm lim}_{500}=3.5 \times 10^{-3}$. 
The corresponding mass limit $M_{500,\rm lim}$ can be obtained from eq.~(\ref{ymrel}) and it depends both on redshift and cosmology.
The predicted counts for the fiducial cosmology $N^*_{\rm th}(z)$ are shown as the solid line in Fig.~\ref{fig:planck_counts}. The total number of clusters between $z=0$ and $z=5$ is 516.3. 
As a term of comparison, we also considered a higher $S/N$ selection ($Y^{\rm lim}_{500}=6.5 \times 10^{-3}$) with no missing redshifts. This yields 201.4 clusters and can be thought of as being similar to the $S/N=7$ cosmological sample used in \cite{Planck20}. 

\paragraph{Survey 2}
The second survey mimics the SPT-SZ complete survey, which will contain around 500 clusters \citep{SPT2013}, and is similar also to that performed by ACT. It has a sky coverage of 2500\,deg$^2$, constant mass limit $M_{500,\rm lim}=7.75 \times 10^{14}\,M_{\odot}$ and no clusters below $z = 0.25$. The predicted counts $N^*_{\rm th}(z)$ for this selection function are shown as a dashed line in Fig.~\ref{fig:planck_counts}; the total numer of clusters is 514.5. This is similar to Survey 1 but with a totally different redshift distribution.

For each survey, we computed the total catalogue counts, $N_{\rm tot}(z)$, directly as the fiducial counts $N^*_{\rm th}(z)$. As pointed out by \cite{sahlen}, using such an ``average catalogue'', as opposed to a particular realization, allows an unbiased recovery of the input cosmological parameters. This would highlight eventual biases due to our procedure for dealing with the missing redshifts. The counts obtained in this way are used as actual catalogue counts, including Poisson error bars. 

\begin{figure}
\includegraphics[width=7.8cm]{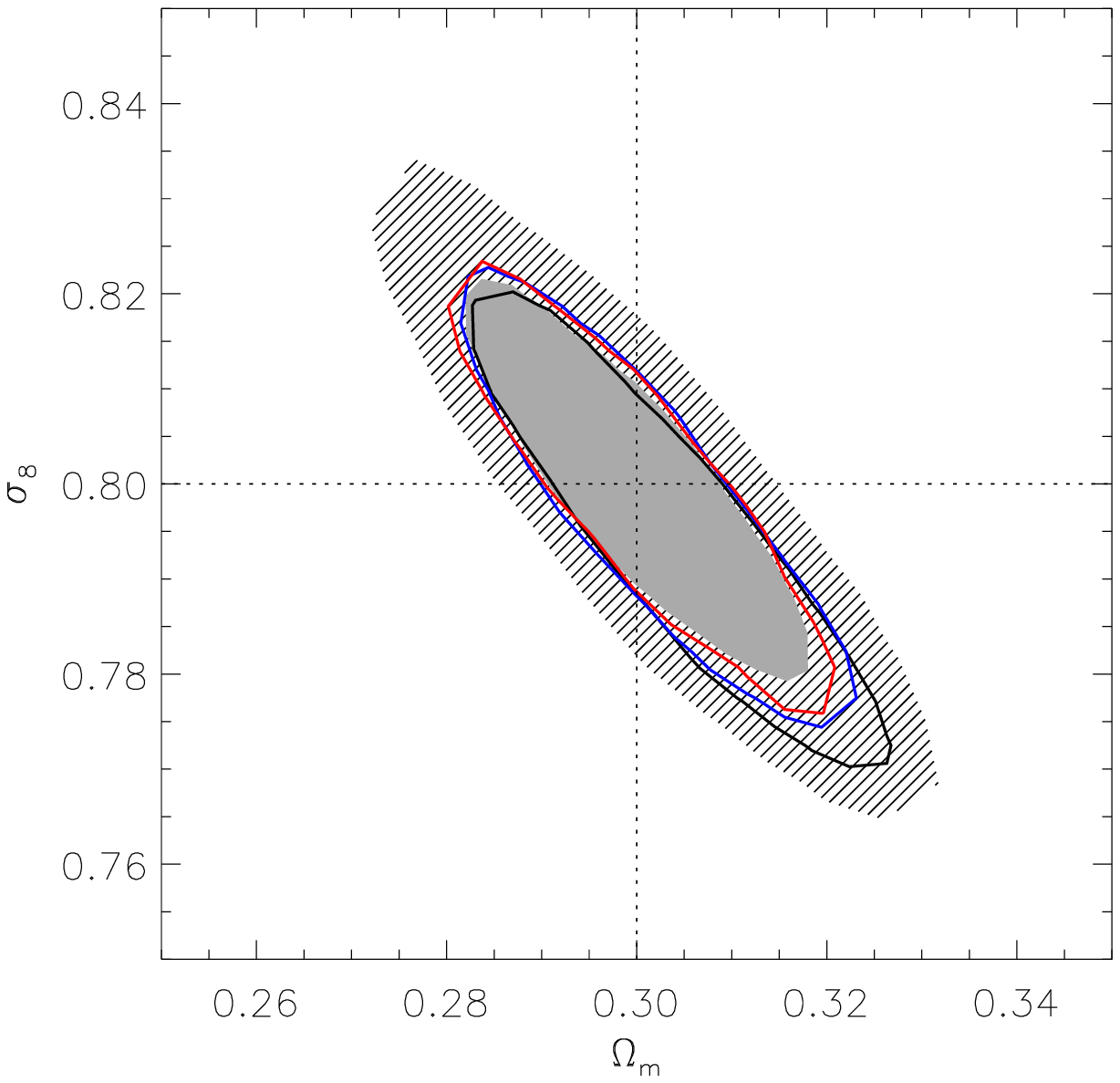}
\includegraphics[width=7.8cm]{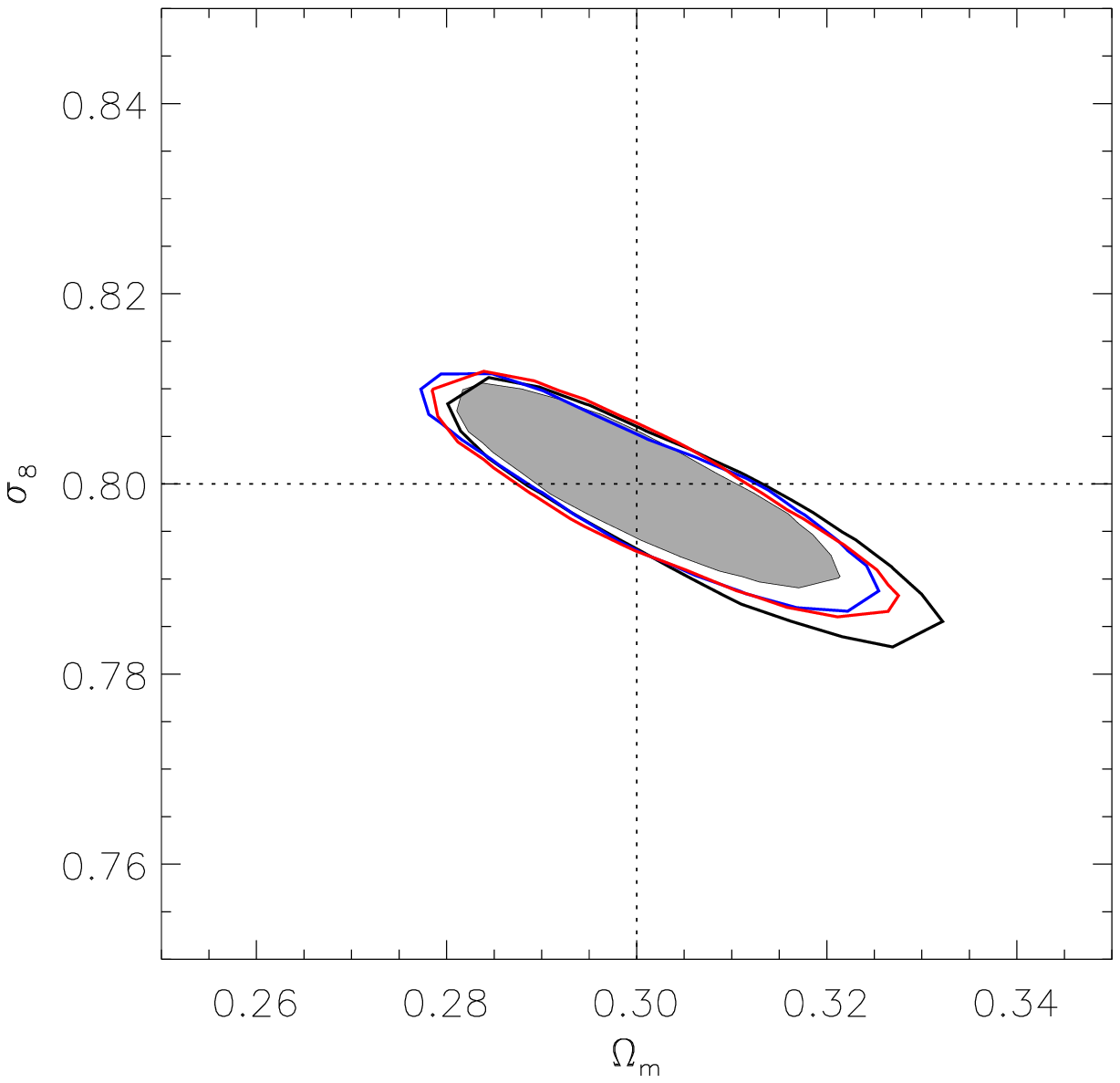}
\caption{Comparison of $\Omega_{\rm m}-\sigma_8$ 68\% CL contours for 0 (grey area); 5\% (red line); 10\% (blue line); and 20\% (black line) missing redshifts for Survey 1 (top) and Survey 2 (bottom) with no redshift selection. The shaded area on the top panel corresponds to the high S/N sample of Survey 1, with no missing redshifts.}
\label{fig:contours_nosel}
\end{figure}
Starting from $N_{\rm tot}(z)$, we  model the distribution of known redshifts, $N_{\rm red}(z)$ as
\begin{equation}
N_{\rm red}(z)=W(z)\,N_{\rm tot}(z).\label{nredz}
\end{equation}
The function $W(z)$ is a generic function of redshift normalised such that
\begin{equation}
\int_{0}^{z_{\rm max}} W(z)N_{\rm tot}(z) dz=N_{\rm red}\label{eq5}
\end{equation}
and represents a selection effect for the known redshifts. In general, $W(z)$ depends on the SZ survey and on the way the redshifts are collected. 

The true distribution of missing redshifts, $N_{\rm miss}(z)$, is clearly $N_{\rm miss}(z)=N_{\rm tot}(z)-N_{\rm red}(z)=[1-W(z)]N_{\rm tot}(z)$. 
However, as this distribution is not known, we use $N_{\rm th}(z)$ to generate realizations of it, $N^j_{\rm miss}(z)$, and compute $N^j_{\rm cat}(z)$ for the likelihood of eqs.~(\ref{cashnew}) and (\ref{cashnew2}) as $N^j_{\rm cat}(z)=N^j_{\rm miss}(z)+N_{\rm red}(z)$.

\section{Results}
In this section we compare the cosmological constraints predicted for $\Omega_{\rm m}$ and $\sigma_8$ for Survey 1 and Survey 2 as a function of the number of missing redshifts. We first consider the case where the missing redshifts are excluded randomly from the catalogues (Sect.~\ref{sec:nosel}), and then consider the impact of selection effects and ways to account for them (Sects.~\ref{sec:selection} and  \ref{sec:modelling}).

\subsection{No selection effects}\label{sec:nosel}
If there is no selection effect for the known redshifts, $W(z)={\rm const}$ and  $N_{\rm red}(z)$ is simply $N_{\rm tot}(z)$ rescaled by $N_{\rm red}/N_{\rm tot}$. As a consequence, the assumption made by our method about the distribution of the missing redshifts is verified.
The cosmological constraints are shown in Fig.~\ref{fig:contours_nosel}. The increase in the 1D error bars with respect to the $N_{\rm miss}=0$ case is moderate (up to 30--35\% for $N_{\rm miss}=0.2\,N_{\rm tot}$). Biases on $\Omega_{\rm m}$ and $\sigma_8$ are much smaller than 1\,$\sigma$ for low $N_{\rm miss}$, and reach 0.2--0.4\,$\sigma$ for $N_{\rm miss}=20\%$. For Survey 1, the results with up to 20\,\% missing redshifs are significantly better than the high $S/N$ case with no missing redshifts, shown as a dashed area in the top panel of Fig.~\ref{fig:contours_nosel}. The latter yields $\Omega_{\rm m}= 0.301\pm0.019$ and $\sigma_8=0.798 \pm0.023$; this is somewhat stronger than the results presented in \cite{Planck20} because in our case we fix all parameters except $\Omega_{\rm m}$ and $\sigma_8$.

\begin{figure}
\includegraphics[width=5.4cm,angle=90]{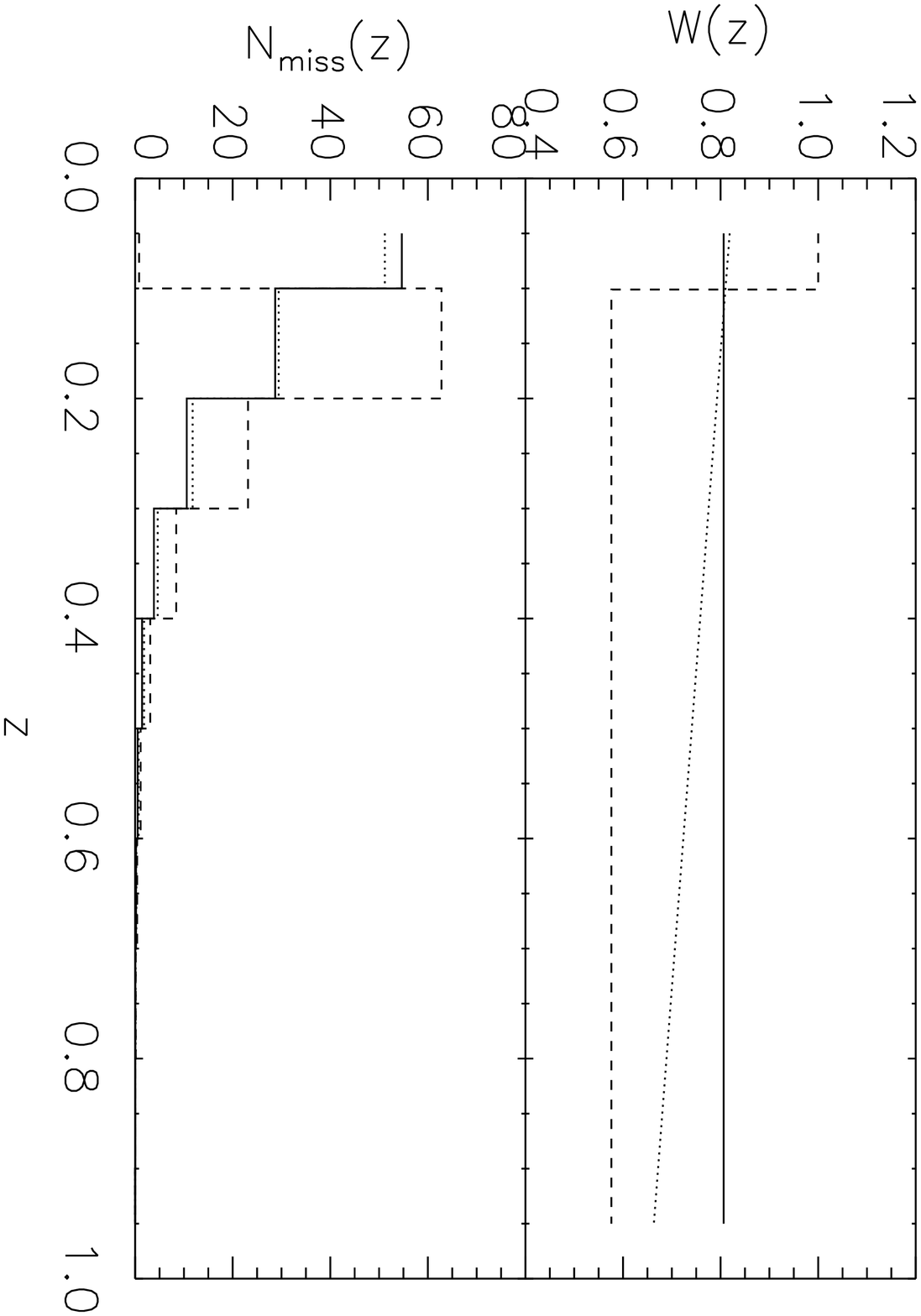}
\includegraphics[width=5.4cm,angle=90]{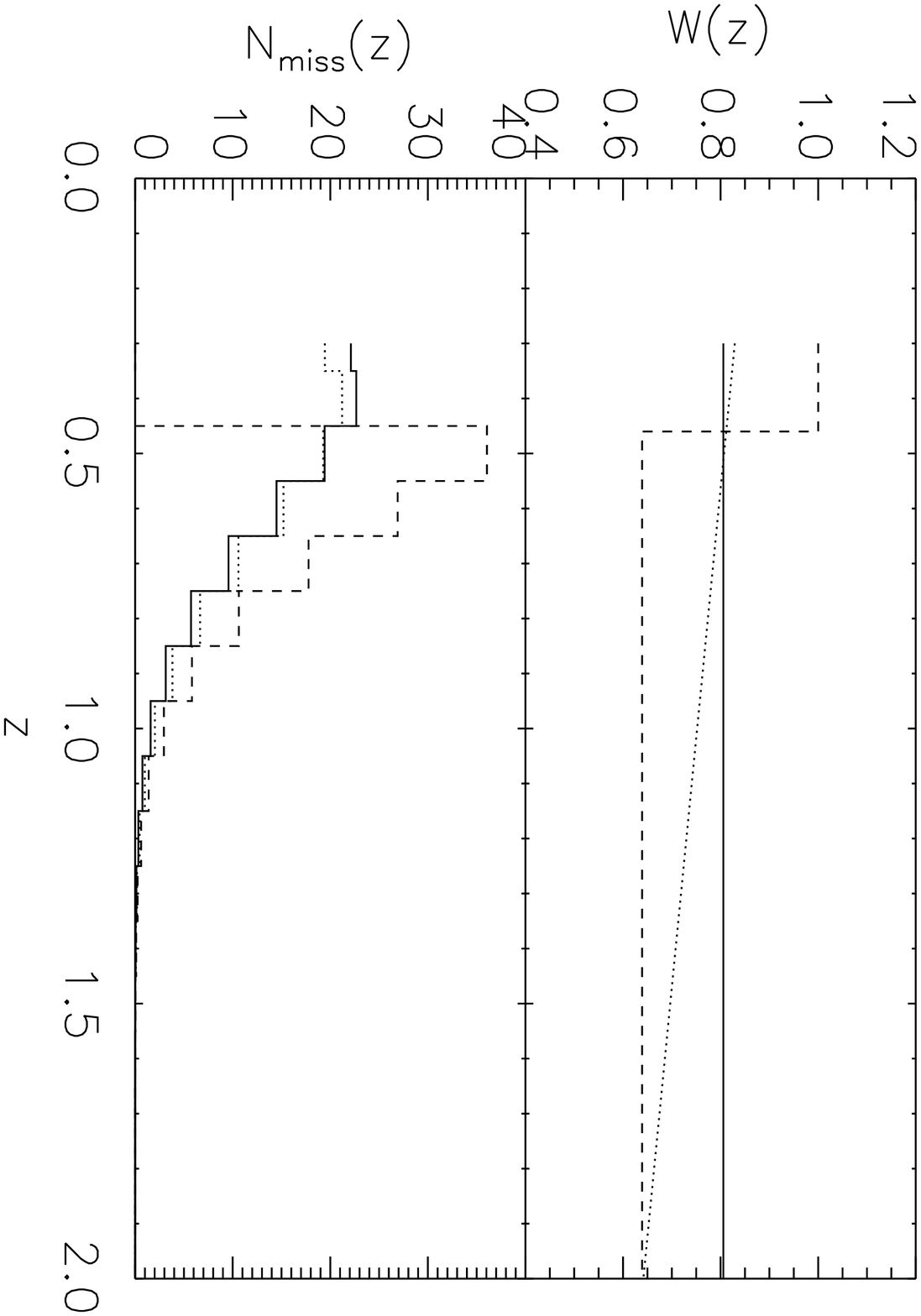}
\caption{Selection functions $W(z)$ and distribution of missing redshifts $N_{\rm miss}(z)$ for $N_{\rm red}/N_{\rm tot}=$80\% (20\% missing redshifts) for Survey 1 on the top panel and Survey 2 on the bottom panel. {\it Solid lines}: no selection effects; {\it dotted lines}: linear selection; {\it dashed lines}: step selection.}
\label{fig:selections}
\end{figure}

\subsection{Mild vs strong selection effects}\label{sec:selection}
We now consider the impact of a non trivial selection $W(z) \neq {\rm const}$ in order to quantify the biases in the cosmological parameters when such an effect is neglected. We considered two opposite situations: a mild selection ($W(z)$ varying slowly with redshift) and a strong selection ($W(z)$ varying rapidly with redshift). Both are intended to be indicative rather than representative. In both cases we adopted decreasing functions of the redshift because low-$z$ clusters are easier to detect and observe.
\begin{figure}
\includegraphics[width=5.4cm,angle=90]{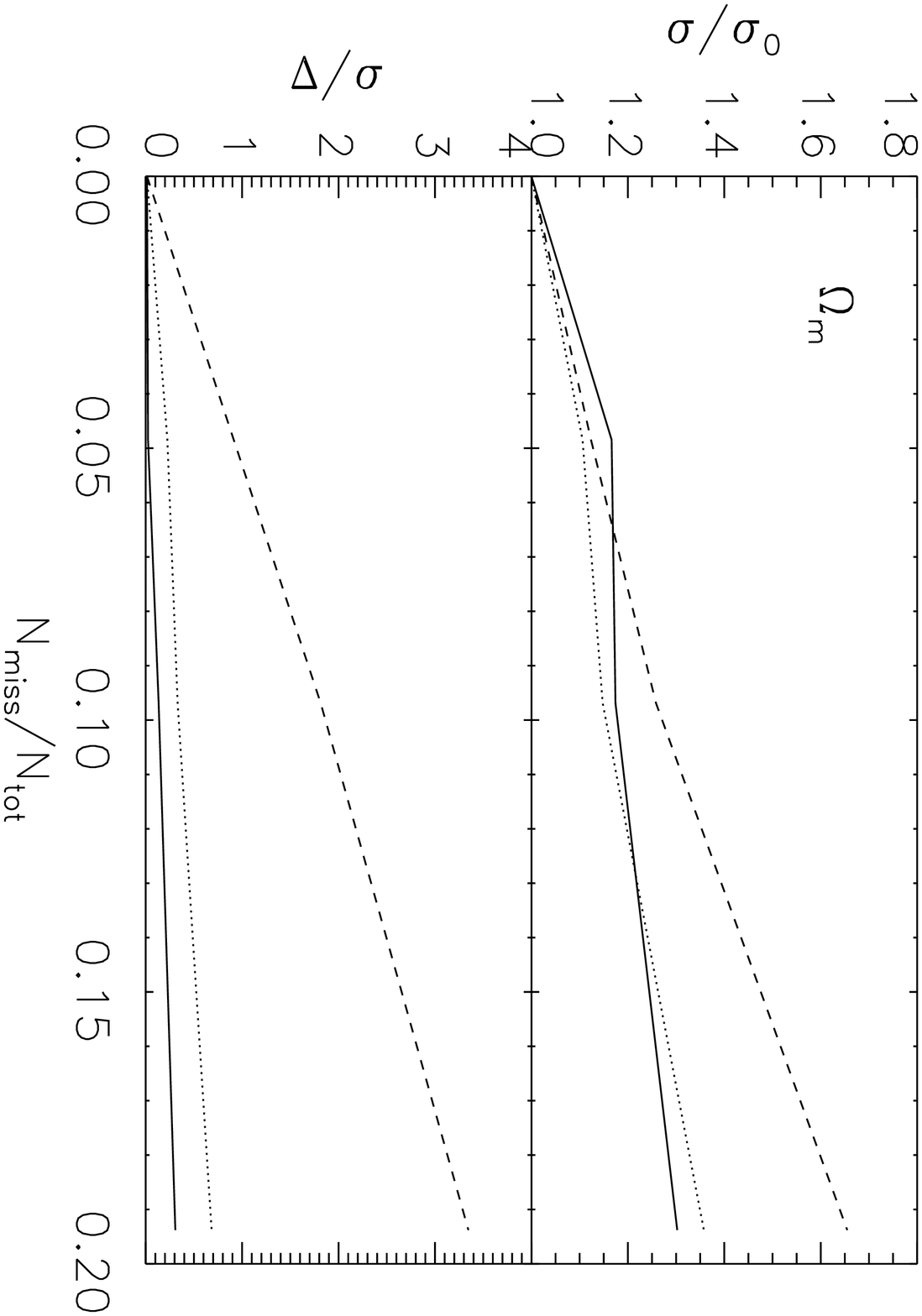}
\includegraphics[width=5.4cm,angle=90]{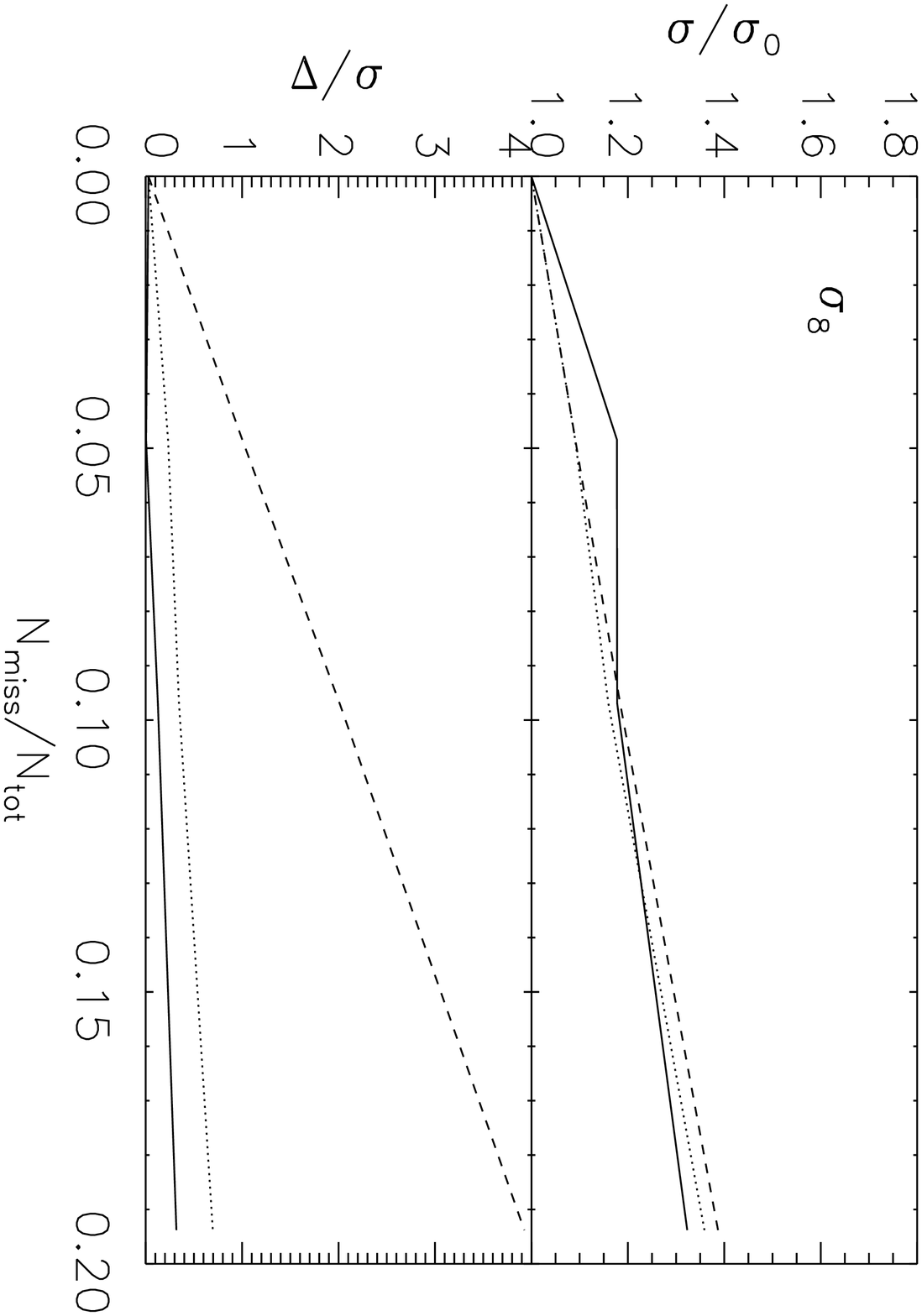}
\caption{68\% CL error bar ($\sigma$) and absolute value of the bias ($\Delta$) as a function of the number of missing redshifts $N_{\rm miss}$ for Survey 1. The error bar is in units of the same quantity with no missing redshifts $\sigma_0$ and the bias is in units of the error bar $\sigma$. Top panels: results for $\Omega_{\rm m}$; bottom panels: results for $\sigma_8$.{\it Solid lines}: no selection effects; {\it dotted lines}: linear selection; {\it dashed lines}: step selection.}
\label{fig:biases}
\end{figure}

For the mild selection we used a linear function of the redshift, while as a strong selection we used a step function ($W(z)=1$ for $z \leq z_0$  and $W(z)=c$ elsewhere). The slope of the linear function and the redshift $z_0$ must give $N_{\rm red}(z) \leq N_{\rm tot}(z)$ for all redshifts. The normalization of the linear function and the amplitude of the step function, $c$, are constrained by eq.~(\ref{eq5}). 
The linear selection is such that the ratio of missing to total clusters in the lowest relevant redshift bin is half that in the highest one. The step selection has no missing redshifts in the first bin for Survey 1 and in the first two bins for Survey 2. In Fig.~\ref{fig:selections} we compare the selection functions and the corresponding $N_{\rm miss}(z)$  distributions for 20\% missing redshifts. We note that the linear selection function modifies the $N_{\rm miss}(z)$ distribution only slightly, while the step selectin function is much stronger. 

\begin{figure}
\includegraphics[width=7.8cm]{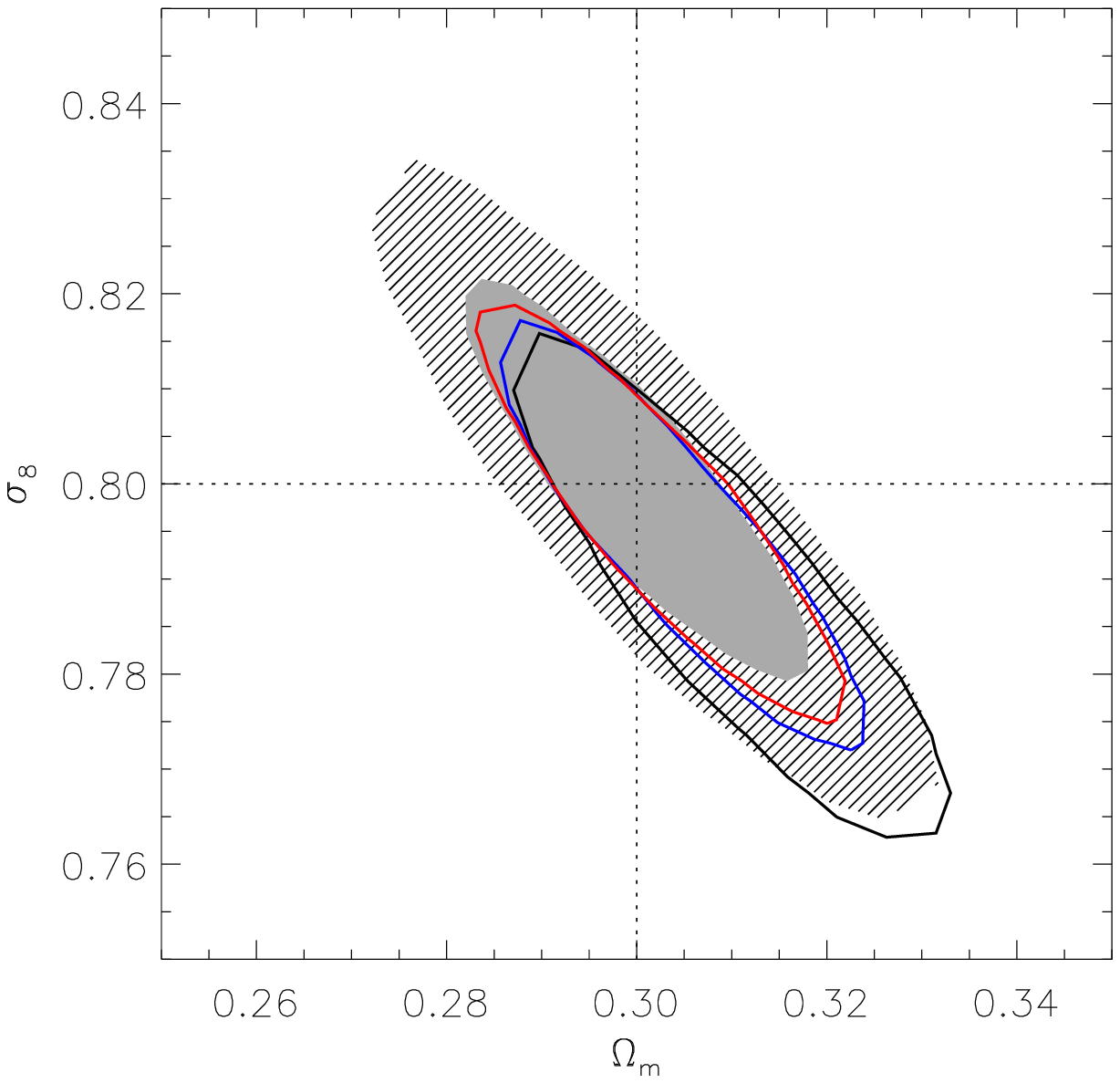}
\includegraphics[width=8cm]{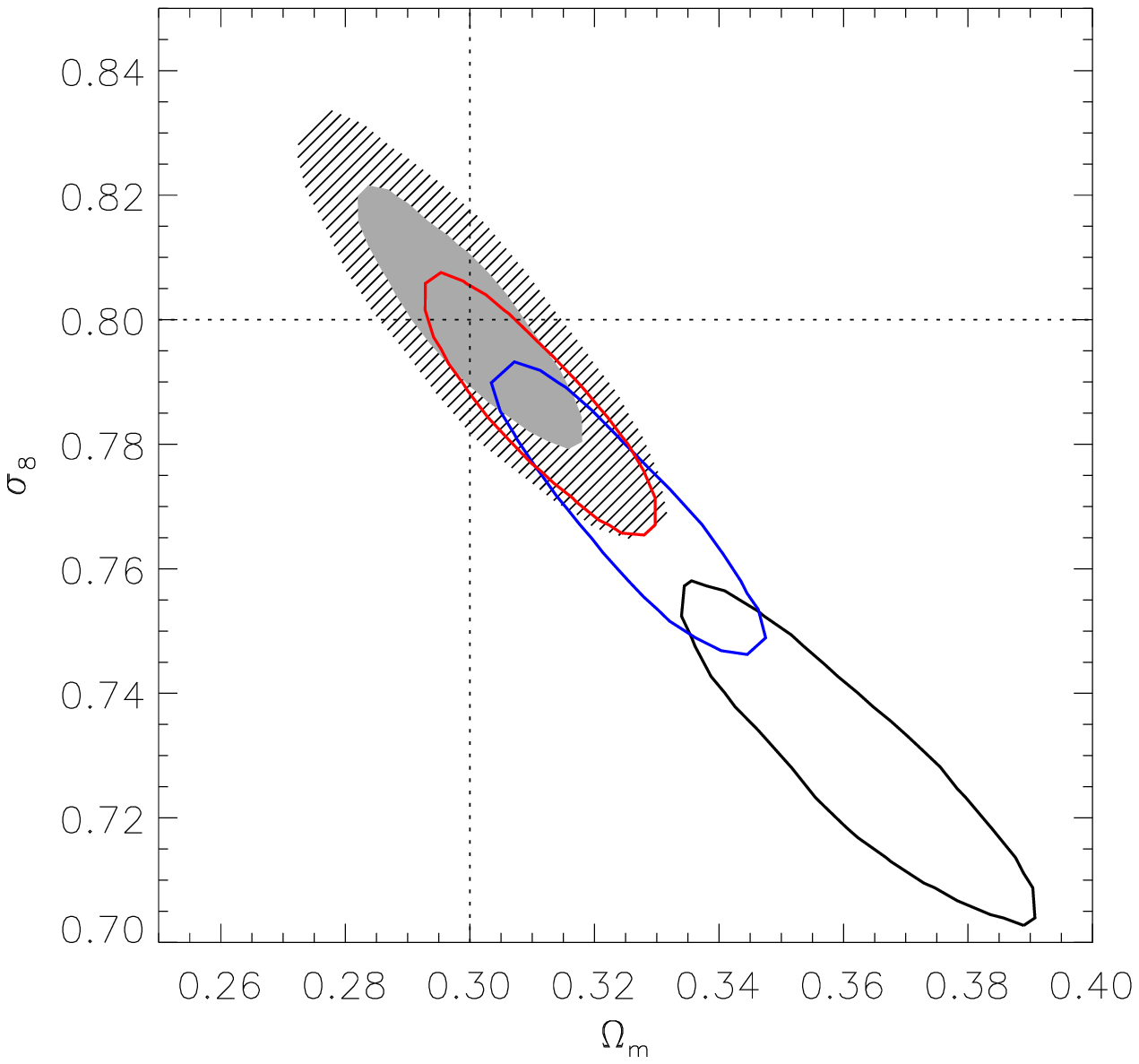}
\caption{Same as Fig.~\ref{fig:contours_nosel} for Survey 1 with linear (top) and step (bottom) selection function not modelled into the analysis.}
\label{fig:contours_sel}
\end{figure}

In Figs. \ref{fig:biases} and \ref{fig:contours_sel} we show the results on the cosmological parameters for Survey 1, those corresponding to Survey 2 being very similar. Fig.~\ref{fig:biases} shows the increase of the 68\% CL error bar on $\Omega_{\rm m}$ and $\sigma_8$ and the absolute value of the bias on the best-fit parameters as a function of the number of missing redshifts. The size of the error bars can either increase or decrease moderately with respect to the case with no selection, depending on the number of missing redshifts and the choice of $W(z)$. Selection effects mainly impact the best-fit value of the cosmological parameters, which will be biased. The bias is moderate ($<1\,\sigma$) for the linear selection function, and much more pronounced (up to 4\,$\sigma$) for the step selection function. 

Fig.~\ref{fig:contours_sel} shows the contours for $\Omega_{\rm m}$ and $\sigma_8$ for Survey 1. Interestingly, the contours shift along the degenerate line. This suggests that the combination of $\Omega_{\rm m}$  and $\sigma_8$ perdendicular to the degenerate line is almost insensitive to the selection effects. We looked for the well measured combination of the form $P=(\Omega_{\rm m}/0.27)^\gamma\sigma_8$ corresponding to the minor axis of the ellipse. We computed the exponent $\gamma$ by performing a principal component analysis of the results with no missing redshifts. We obtained $\gamma=0.45$ for Survey 1 and  $\gamma=0.20$ for Survey 2. The results in terms of this derived parameter $P$ are indeed much more robust against the systematic effects (the bias being at worse $1.4\,\sigma$ instead of $4\,\sigma$). This suggests that $P$ is a measure of the total number of clusters and the other principal component depends on the redshift distribution. 

\subsection{Modelling selection effects} \label{sec:modelling}
The results of Sec.~\ref{sec:selection} support the conclusion that strong selection effects must be accounted for to avoid significant biases on $\Omega_{\rm m}$ and $\sigma_8$. If $W(z)$ is known, this can be achieved by drawing the missing redshifts from a probability distribution given by $[1-W(z)] \,N_{\rm th} (z)$ instead of $N_{\rm th} (z)$. 
We checked that, once modelled correctly into the analysis, the step selection function yields unbiased results, comparable with those shown in Fig.\ref{fig:contours_nosel}. 

When dealing with real surveys, the redshift accumulation strategy could provide a prior knowledge of $W(z)$. Alternatively, the selection function could be estimated from the data, for example by measuring all the redshifts for a ``deep'' region of the survey and comparing the distribution $N_{\rm tot}(z)$ for this region with $N_{\rm red}(z)$ for the rest of the survey. Assessing the impact of errors on $W(z)$ on the cosmological parameters is beyond the scope of this paper since it would be very survey specific. However, we suggest that a reasonable estimate of $W(z)$ would only leave mild residual selection effects, which would then have a limited impact on the results.
 
\section{Conclusions}
We have presented a statistical method for dealing with incomplete redshift coverage of SZ cluster catalogues during the estimation of cosmological parameters. We tested it on simulated data which mimic those of SPT/ACT and {\it Planck}. 
For a number of missing redshifts up to 20\% and in the absence of systematic effects in the accumulation of redshifts, the recovery of the cosmological parameters is almost unbiased and there is only a mild increase in the 1D error bars (up to 35\,\%). 

Selection effects for the known redshifts, if not accounted for, can cause significant offsets in the cosmological parameters. The contours shift along the degenerate line, allowing a robust computation of the parameter perpendicular to it. We showed how known systematic effects can be modelled into the analysis to avoid biases in the cosmological parameters.

\section{Acknowledgements}
AB and MLB acknowledge support from the European Research Council under the EC FP7 grant number 280127. MLB also acknowledges support from an STFC Advanced/Halliday fellowship. We thank the anonymous referee for useful comments. We thank A. Liddle, J.-B. Melin, and the members of the Planck Collaboration working on the cosmology with SZ project, for useful inputs and discussion.


\end{document}